\documentclass[useAMS]{mn2e}
\usepackage{epsfig}
\title[A new globular cluster black hole]{A new globular cluster black
hole in NGC~4472}

\author[Maccarone et al.]{Thomas J. Maccarone\\ School of Physics and
Astronomy, University of Southampton, Hampshire SO17 1BJ,United
Kingdom\\ \newauthor Arunav Kundu\\ Eureka Scientific, 2452 Delmer
Street Suite 100, Oakland, CA 94602-3017, USA\\ \newauthor Stephen
E. Zepf\\ Department of Physics and Astronomy, Michigan State
University, East Lansing, MI 48824, USA\\ \newauthor Katherine
L. Rhode\\ Department of Astronomy, Indiana University,727 East 3rd
Street, Bloomington, IN 47405-7105, USA }

\begin{document}
\def\ltsim{\mathrel{\rlap{\lower 3pt\hbox{$\sim$}}
        \raise 2.0pt\hbox{$<$}}}
\def\gtsim{\mathrel{\rlap{\lower 3pt\hbox{$\sim$}}
        \raise 2.0pt\hbox{$>$}}}

\date{}

\pagerange{\pageref{firstpage}--\pageref{lastpage}} \pubyear{}

\maketitle

\label{firstpage}
\begin{abstract}
We discuss CXOU~1229410+075744, a new black hole candidate in a
globular cluster in the elliptical galaxy NGC~4472.  By comparing two
Chandra observations of the galaxy, we find a source that varies by at
least a factor of 4, and has a peak luminosity of at least
$2\times10^{39}$ ergs/sec.  As such, the source varies by
significantly more than the Eddington luminosity for a single neutron
star, and is a strong candidate for being a globular cluster black
hole.  The source's X-ray spectrum also evolves in a manner consistent
with what would be expected from a single accreting stellar mass black
hole.  We consider the properties of the host cluster of this source
and the six other strong black hole X-ray binary candidates, and find
that there is suggestive evidence that black hole X-ray binary
formation is favored in bright and metal rich clusters, just as is the
case for bright X-ray sources in general.  

\end{abstract}


\section{Introduction}

In recent years there has been a revolution in our understanding of
black holes in globular clusters.  Until the launches of Chandra and
XMM-Newton, it had become widely, if not universally, accepted lore
that stellar mass black holes could not exist in significant numbers
in globular clusters.  Spitzer (1969) considered a hypothetical star
cluster whose stars had only two possible masses.  He found that given
a large mass ratio between the two components and a significant
fraction of the stars in the more massive component, an instability
would lead to severe mass segregation, such that the heavier stars
would form a sub-cluster which would not be affected by the outer main
cluster.  This sub-cluster can then evaporate itself dynamically on a
timescale much less than the Hubble time.  The combination of
theoretical work, plus the lack of observational evidence for stellar
mass black holes in Milky Way globular clusters led to the suggestion
that globular clusters should not, and did not, contain substantial
populations of black holes.  More recent theoretical work, however,
has shown that substantial fractions of black holes can be retained in
globular clusters (Mackey et al. 2007; Moody \& Sigurdsson 2009).  The
finding that some black holes are retained should not be surprising
given that the Spitzer instability requires that the ratio of the
total mass in the heavy component to the total mass in the light
component exceeds a critical value that is a function of the ratio of
the masses; i.e. once a certain number of black holes are ejected, the
Spitzer instability criterion is no longer satisfied.

In the early Chandra observations of elliptical galaxies, many sources
were found with luminosities exceeding the Eddington luminosity of a
single neutron star (e.g. Sarazin et al. 2000; Angelini et al 2001).
Nevertheless, there remained in those cases the possibility that the
X-ray sources were not single black holes, but rather superpositions
of several bright neutron stars; variability is needed to distinguish
between the two possibilities (Kalogera et al. 2004).  The discovery
of strong variability from the $4.5\times10^{39}$ ergs/sec source,
XMMU~122939.7+075333 in NGC~4472, by Maccarone et al. (2007) confirmed
that, in at least one case, a source could not be explained as a
superposition of Eddington-limited neutron stars, and thus had to be a
black hole, or a highly beamed neutron star.  The more recent
discoveries of strong, broad [O III] emission lines (Zepf et
al. 2007;2008) from this system ruled out the possibilities of
substantial beaming (since the [O III] forbidden line emission, must
come from an optically thin and hence isotropically emitting region --
see e.g Gnedin et al. 2009).  This result thus strongly supports a
scenario with a $L_X\gtsim L_{EDD}$ system, in order to drive the
strong outflows implied by the optical emission lines.  The system is
therefore likely to host a a stellar mass black hole accretor, rather
than an intermediate mass black hole.

Recently, at least two other cases for black holes in globular
clusters have been identified (Brassington et al. 2010; Shih et al.,
submitted to ApJ).  Brassington et al. (2010) found evidence for a
source in NGC~3379 near the Eddington luminosity for a stellar mass
black hole, which varied by about 30\% (with a $\Delta L$ larger than
the Eddington luminosity for a neutron star), and with X-ray spectra
consistent with the expectations for a stellar mass black hole in that
luminosity range.  Shih et al. (2010) report a more complicated
pattern of source behavior -- a source in NGC~1399 which has several
epochs near or slightly above $L_X=10^{39}$ ergs/sec, and variability
of factors of several within an observation and a factor of $\sim 100$
on timescales of years.  In neither case is there any evidence put
forth to support the idea that these might be intermediate mass black
holes.  A different type of case has also been made for another black
hole in a globular cluster in NGC~1399 on the basis of a high ($\sim
4\times10^{39}$ ergs/sec) X-ray luminosity and strong [O III] and [N
II] emission lines (Irwin et al. 2010).  While Irwin et al. (2010)
argued for an interpretation favoring an intermediate mass black hole
based on the soft X-ray spectrum and high luminosity, we note that
there are alternative accretion disc models in which such a soft
spectrum at a high luminosity can result from a mildly super-Eddington
accretion flow (e.g. Soria et al. 2007; Gladstone et al. 2009).

Recently, there has also been a discovery of a very strong
intermediate mass black hole candidate in the galaxy ESO243-49, which
has an extremely high luminosity, of about $10^{42}$ ergs/sec (Farrell
et al. 2009).  The source also shows spectral state transitions at
about $3\times10^{41}$ ergs/sec (Godet et al. 2009), which imply a
black hole mass of about $10^4 M_\odot$ if they occur at the typical
2\% of the Eddington luminosity (Maccarone 2003).  Recently, the
source has been found to have an optical counterpart whose flux is
consistent with expectations of a bright globular cluster in
ESO~243-49 (Soria et al. 2010), but the identification of the optical
counterpart as a globular cluster remains highly insecure for the time
being.  We thus mention this source for completeness, but do not
include it in any of the analysis of the properties of globular
clusters hosting black hole candidates.  In this paper, we present
evidence for a fourth variability-confirmed globular cluster black
hole (and fifth strong candidate), and discuss the properties of the
global sample of strong globular cluster black hole candidates found
to date.

\section{Data analysis}

Chandra has observed NGC~4472 three times, and there exists a fourth
field not aimed at NGC~4472 itself, but which includes our source of
interest.  Two of the observations of NGC~4472 were taken in 2000, one
with ACIS-I on 19 March 2000, and the other with ACIS-S on 12 June
2000.  The most recent was taken on 27 February 2010.  The observation
in which our source appears, but which was not an observation of
NGC~4472 was taken 23 February 2008 as part of the AMUSE-Virgo project
(see e.g. Gallo et al. 2008).  We produced a catalog of X-ray sources
using WAVDETECT (Freeman et al. 2002) after processing the data to
remove high background time intervals.

We then used TOPCAT to combine the catalog of X-ray sources produced
from this observation with both the catalog from the previous Chandra
observations of NGC~4472 discussed in Kundu et al. (2002) and
published in Maccarone et al. (2003 -- MKZ03) and the optical catalog
of globular clusters published in MKZ03.  We searched the list of
sources that matched globular clusters in the optical catalog for
sources with luminosities above $5\times10^{38}$ ergs/sec.  We then
checked these sources for variability between the two epochs of long
Chandra integrations.  We identified one new variable source
associated with globular cluster 28 of the catalog of MKZ03.  This
source, located at 12h29m41.0s +7$^\circ$57'44'', about 2.7' from the
center of NGC~4472, was detected in 2001 by Chandra, but we did not
report on its in KMZ02 or MKZ03 because the source fell on the ACIS-S2
chip, and those papers reported only sources on ACIS-S3. 

We find a separation of 0.5'' between the X-ray position from the 2010
observations and the optical position from MKZ03.  We test the chance
superposition probability by making shifts to the positions of the
X-ray sources by small amounts, the determining the number of matches
closer than 0.65'', the separation beyond which Kundu et al. (2002)
found there were few additional matches, indicating that most of the
additional matches were likely to be chance superpositions.  We
restrict our estimate of the number of false matches to the region
between 2.2 and 3.2 arcminutes from the center of the galaxy in order
to ensure that the space density of optical clusters in the region
searched is similar to that around CXOU~1229410+075744, which is
located 2.7' from the center of NGC~4472.  Within this region, there
are 23 X-ray sources, and 343 optical clusters.  Formally, the region
subtends a solid angle of about 61000 square arcseconds, but about 1/5
of the region was not covered with the HST observations used for the
comparisons.  Regions matching an optical cluster within 0.65'' thus
subtend about 1\% of the region.  We thus estimate a 1\% chance
superposition probability for a given X-ray source with an optical
globular cluster (even considering a 1'' matching radius would give a
less than 3\% probabilty of a chance superposition).  The particular
cluster with which this X-ray source matches is among the brightest
10\% of the clusters in NGC~4472, which can be taken as additional
support of the idea that the match is genuine, albeit in an {\it a
posteriori} manner that is difficult to quantify.

\subsection{Deep Chandra observations}
Having identified a strong black hole candidate, we then produced
X-ray spectra of the source using the specextract tool and light
curves of the source using the dmextract tool on the two deep ACIS-S
observations.  The spectra were extracted with an 8 pixel radius, and
an offset background region of 20 pixel radius.  The channels in the
source spectrum were then binned in groups of 20 or more photons.
Spectral fitting was done in XSPEC 12.5.1 (see Arnaud 1996 for a
description of an earlier version of XSPEC)\footnote{For a more
up-to-date description of XSPEC, see
http://heasarc.gsfc.nasa.gov/docs/xanadu/xspec/}.  Rebinned channels
including photons below 0.5 keV and above 8 keV are ignored, because
the response matrix for Chandra is better calibrated within this
energy range than outside of it.  We report the 1$\sigma$ errors on
source parameters.

In observation 321, taken 12 June 2000, the cluster's X-ray spectrum
was well fitted ($\chi^2/\nu=4.74/4$) with a disk blackbody model with
foreground neutral hydrogen column density of $1.6\times10^{20}$
cm$^{-2}$, $kT = 0.88^{+0.16}_{-0.13}$ keV and normalization in XSPEC
of 2.0$^{+1.6}_{-0.9} \times10^{-3}$, corresponding to a best-fitting
inner disk radius of 72 km assuming a face-on disk and no spectral
hardening correction (Mitsuda et al. 1984).  The source flux from
0.5-8.0 keV is 2.2$^{+0.1}_{-0.9}\times10^{-14}$ ergs/cm$^2$/sec.  The
best fitting value corresponds to a luminosity of $6.5\times10^{38}$
ergs/sec using a distance of 16 Mpc to NGC~4472, based on the distance
to the Virgo Cluster in which it is contained (Macri et al. 1999).

\begin{figure*}
\psfig{file=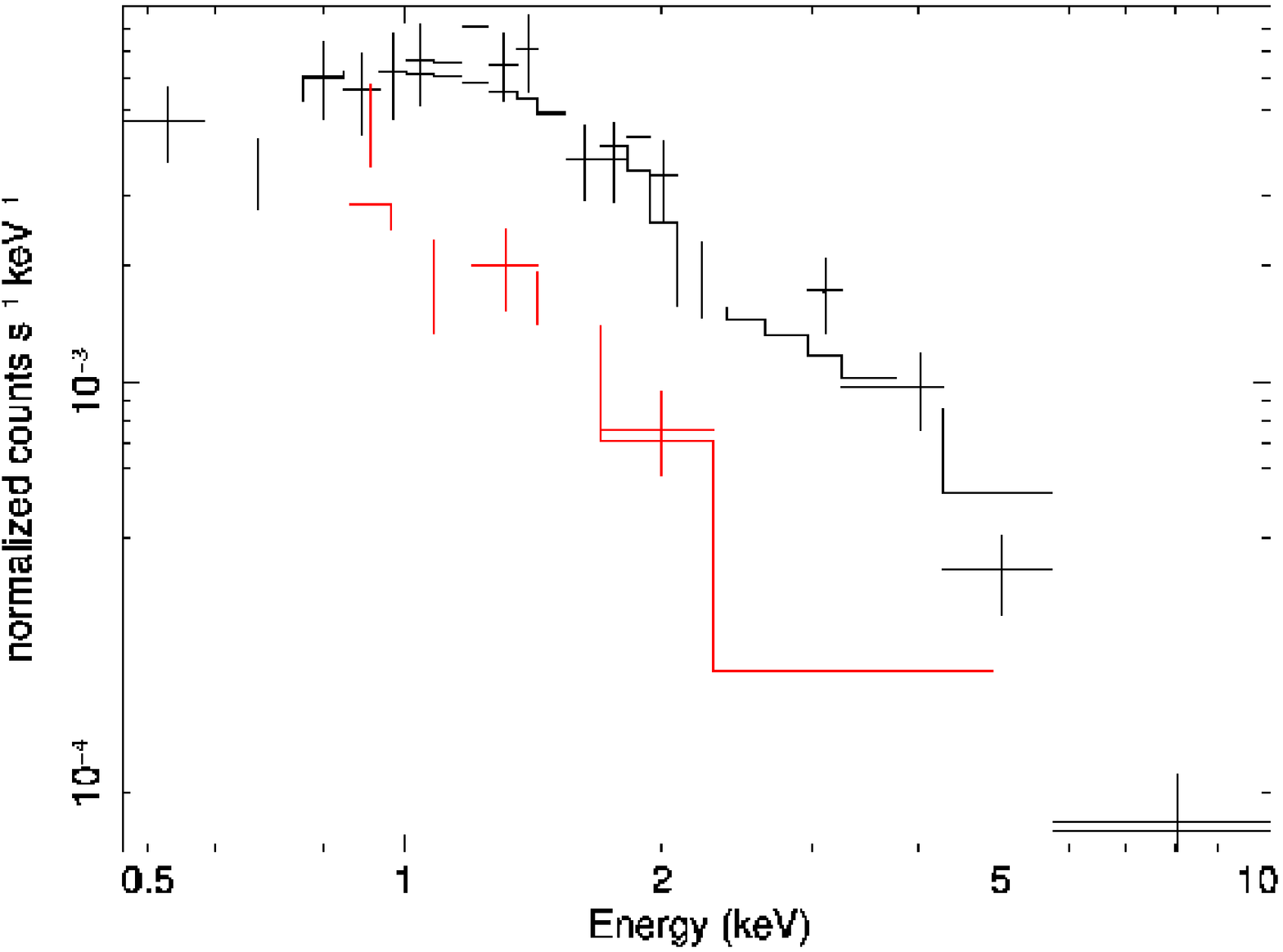, width=2.6 in, angle=-90}\psfig{file=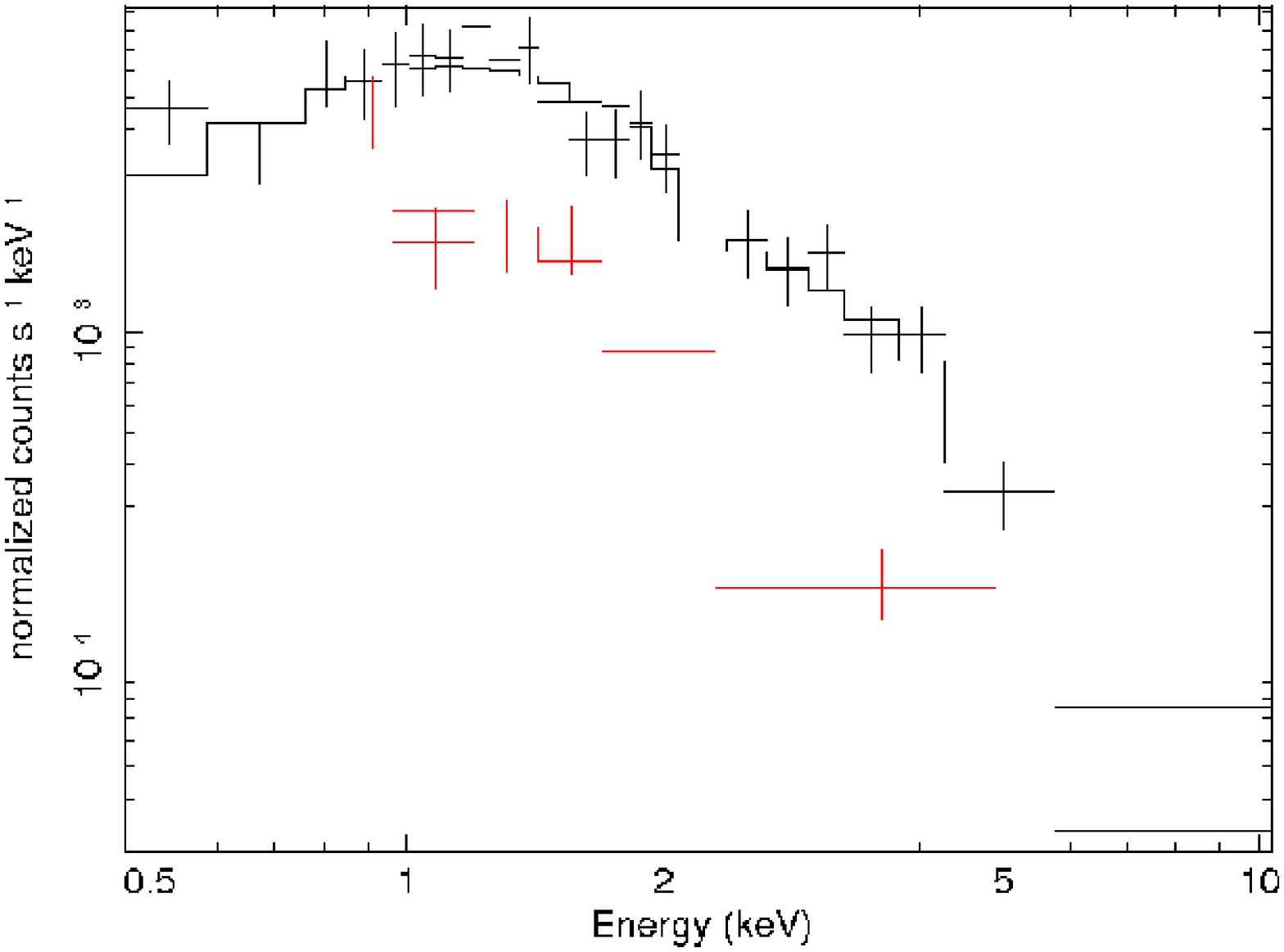, width=2.6 in, angle=-90}
\caption{The spectra from the deep Chandra observations, showing that
the strong variability is unambigously detected, and that the $\Delta
L$ is significantly larger than the Eddington luminosity of a single
neutron star.  The power law model plotted over the data in the figure
on the left, and the disk blackbody model plotted over the data in the
figure on the right.  Observation 11274 is the upper curve in both
cases.}
\label{spectra}
\end{figure*}

In observation 11274, taken on 27 February 2010, the source spectrum
was well fitted (i.e. $\chi^2/\nu$=10.5/19) with $kT =
1.52^{+0.16}_{-0.13}$ keV and normalization in XSPEC of
9.2$^{+3.3}_{-2.5}\times10^{-4}$, corresponding to a best fitting
inner disk radius of 34 km assuming a face-on disk and no spectral
hardening correction.  The source flux from 0.5-8.0 keV is
9.2$^{+0.2}_{-1.4}\times10^{-14}$ ergs/cm$^2$/sec.  The best fitting
value corresponds to a luminosity of $2.7\times10^{39}$ ergs/sec. The
temperatures of the disk are thus different at nearly the $3\sigma$
level, and both spectra are consistent with standard phenomenology
that the inner disk radius will vary little in high/soft states.  The
luminosity difference is significant at more than $4\sigma$, and the
$3\sigma$ lower limit to the luminosity difference is about
$5\times10^{38}$ ergs/sec, above the Eddington luminosity for a single
neutron star.  The Chandra spectra are shown in figure \ref{spectra}.

We have also tried to fit power law models to the data.  For
observation 321, a power law model with the foreground $N_H$ gives an
acceptable fit, with $\chi^2/\nu=4.76/4$, with a power law index of
$2.00\pm^{+0.24}_{-0.23}$, and the 1-$\sigma$ confidence interval for
the flux ranging from 2.3-3.4$\times10^{-14}$ ergs/sec/cm$^2$.  The
spectral shape is thus marginally consistent with expectations for a
low/hard state, but the luminosity is well above the few percent of
the Eddington luminosity in which hard states are typically found
(Maccarone 2003).  We thus favor the diskbb model fit as providing
parameter values more likely to be indicative of the real physical
state of the system, but we do note that the spectral fits do not
distinguish between the two scenarios.  For observation 11274, a power
law model with the foreground $N_H$ is formally a good fit, with
$\chi^2/\nu=23/19$, but with a spectral index of $1.30\pm0.07$,
considerably harder than is ever seen in a low hard state from a
Galactic black hole X-ray binary.  Since in the former case, the power
law model provides a poor fit to the data, and in the latter case, the
best fitting value of the power law index lies outside the range
expected from phenomenology, there is a strong case to be made that
the data are genuinely better explained with a strong thermal
component than a pure power law spectrum.

\subsection{Shallow Chandra observations}
The two other Chandra observations of this field of view, observation
322 (made 19 March 2000), and observation 8095 (made 23 February
2008), have much shorter integration times.  For CXOU~1229410+075744,
observation 322 yields 40 counts in 10 kiloseconds with ACIS-I, and
observation 8095 yields 24 counts in 5 kiloseconds with ACIS-S. Both
observations yield flux levels of $\sim$5 $\times10^{38}$
ergs/sec/cm$^2$.  Because there are not enough counts for detailed
spectral fitting, there is a considerable uncertainty on the
counts-to-energy conversion.  The Poisson errors are also substantial.
As a result, it is difficult to determine whether the flux levels
during the two short observations were higher or lower than those
during the longer observations.  The X-ray detections and upper limits
are summarized in Table \ref{xrays}.

\subsection{XMM-Newton observations}
Two deep XMM-Newton observations of this source have been made as
well.  However, this source is close enough to the center of NGC~4472
that the diffuse gas emission significantly affects XMM's sensitivity.
The 2XMM catalog (Watson et al. 2009) reports a source 8'' away from
CXOU~1229410+075744, with a positional error of 4.47'' at
12h29m40.49s, $+7^{\circ}57$m47.1s on 5 June 2002.  The source is
given a quality flag (the SUM\_FLAG parameter value) of 4, indicating
that it is located within a region where spurious detections are
likely, and that the source itself may be a spurious source.
Formally, $4\sigma$ upper limits can be obtained from the FLIX tool,
\footnote{http://ledas-www.star.le.ac.uk/flix/flix.html} using data
corresponding to the 2XMMi-DR3 data release.  FLIX finds that the
source was no brighter than about $7\times10^{38}$ ergs/sec on 5 June
2002, and $2\times10^{38}$ ergs/sec on 1 January 2004.

While the upper limits from FLIX appear to indicate that the source
faded sometime after 2001, and re-brightened sometime between 2004 and
2008, we have also looked at the aperture photometry from the FLIX
tool.  We set the extraction region to 5'', in order to limit the
effects of confusion from nearby gas emission and other point sources.
We find that in the obseravations made on 1 January 2004, all three
XMM instruments show a flux more than $3.8\sigma$ above background in
the 0.2-12 keV band, with the most sensitive PN detection above
$5\sigma$. The flux within 5'' is $1.9\pm0.3\times10^{-14}$ ergs/sec
in the EPIC-PN, $2.1\pm0.6\times10^{-14}$ ergs/sec in MOS1, and
$3.0\pm0.6\times10^{-14}$ ergs/sec in MOS2.  The encircled energy
fraction at 5'' is about 40\%
\footnote{http://xmm.esa.int/external/xmm\_user\_support/documentation/uhb/node17.html}.
Taking the aperture photometry at face value, we estimate that the
source was at about $1-2\times10^{39}$ ergs/sec.  The aperture
photometry from FLIX for the 5 June 2002 observation gives a flux
level similar to that in the 1 January 2004 observation, but the
source was only about $2\sigma$ above the background on 5 June 2002.
We tentatively trust the aperture photometry results, in part because
they indicate a more physically likely scenario - that the source did
not change sharply in luminosity twice in less than a decade - but
prefer a cautious approach regarding any conclusion sensitive to the
results of the XMM analysis.

\begin{table}
\begin{tabular}{lcccc}
\hline 
Date & Observatory & Exposure & $L_X$ (ergs/sec)\\
\hline 
19 March 2000 & Chandra ACIS-I & 10 ksec & $\sim5\times10^{38}$\\
12 June 2000 & Chandra ACIS-S& 40 ksec & $6.5\times10^{38}$ \\
23 February 2008 & Chandra ACIS-S& 5 ksec& $\sim5\times10^{38}$\\
27 February 2010 & Chandra ACIS-S& 40 ksec&  $2.7\times10^{39}$\\ 
\end{tabular}
\caption{A summary of the Chandra observations of
CXOU~J1229410+075744.  Because XMM's angular resolution causes
problems for sources this close to the center of the galaxy, and the
results of the XMM analysis are ambiguous, we present the XMM results
only in the text and not in the table.  The two luminosities in the
shorter Chandra observations are uncertain by a factor of $\sim2$
because the integrations were not long enough to allow for reliable
spectral fitting, leaving the counts-to-energy conversion factor
uncertain.}

\label{xrays}
\end{table}

\subsection{Optical properties of the cluster}
The cluster has $V$=21.77 and $V-I$ = 1.34 (MKZ03).  The object is
spectroscopically confirmed to be a globular cluster (Zepf et
al. 2000).  Using the color-metallicity relation of Smits et
al. (2006), the V-I color corresponds to a metallicity of [Fe/H]=+0.4.
The cluster is thus at the metal rich end of the distribution of
clusters, but is probably not quite as metal rich as suggested by the
linear interpolation from Smits et al. (2006).  We can also look at
the ground-based optical photometry on this cluster.  Imaging in $BVR$
with the Mosaic camera on the NOAO-4m yielded $V=21.55, B-V=0.85,
V-R=0.59$ for the cluster (Rhode \& Zepf 2001).  found this cluster to
have $V=21.55, B-V=0.85, V-R=0.59$.  The minor difference in $V$ is
likely mostly due to some combination of the slightly different
bandpass for the HST $V$ band filter than for ground-based Johnson
filters and small differences in the aperture corrections needed for
ground-based versus space-based photometry of mildly extended objects
(e.g. Kundu 2008).  The colors in both the HST data and ground-based
data make this cluster among the 10\% reddest clusters in NGC~4472.
At the present time, there are not HST data deep enough to estimate
either the core radius or the stellar interaction rate for this
cluster.

\section{Discussion}
There are now five strong candidate globular cluster black holes.  For
three of these sources, the black hole nature has been confirmed by a
change in luminosity by an amount in excess of the Eddington limit for
a neutron star(M07; Brassington et al. 2010; Shih et al. 2010), while
for the fifth, the source luminosity is in excess of $10^{39}$
ergs/sec, and the source shows peculiar optical emission lines (Irwin
et al. 2010).  It is interesting to note that two of the sources
showed strong variability within a Chandra observation (Maccarone et
al. 2007; Shih et al. 2010), while this source does not show such
variability.  This source, and the black hole candidate in NGC~3379,
show behavior typical of stellar mass black hole X-ray binaries in the
Galaxy, with no strong variability on timescales of hours, and with
spectral shapes consistent with hot (i.e. $K_BT\sim1$ keV) disk
blackbodies.  If CXOU~1229410+075744 continues to rise in luminosity,
it may eventually make a transition to an ``ultraluminous'' state,
with a softer spectrum, and it would then be interesting to search its
host cluster for optical emission lines.

The properties of the host clusters are listed in Table
\ref{clusters}.  We have checked what fraction of clusters in each
galaxy are redder than the clusters containing strong black hole
candidates, using the catalog used to estimate the colours of the
cluster in question.  Two of the clusters, NGC~4472B and NGC~1399B,
are in the reddest 10\% of clusters in their host galaxies (these
clusters are bluer than only 72/928 and 39/575 of the clusters in the
catalogs of MKZ03 and Dirsch et al. 2004).  The NGC~3379 cluster which
contains a variability-confirmed black hole candidate is bluer than
only 22/61 clusters, and the cluster hosting NGC~1399A is bluer than
only 129/554 clusters in the catalog of KMZ07.  Only NGC~4472A is in a
cluster bluer than the mean value.

A Kolomogorov-Smirnov test can be performed to determine whether the
suggestive evidence that the metal rich clusters are more likely to
contain black holes than the metal poor clusters is statistically
significant.  We take the fraction of clusters bluer than the
BH-hosting clusters for each galaxy, and make a cumulative
distribution of them, then compare with a uniform distribution.  The
largest difference between the two distributions is 0.44, with 5
objects, giving a null hypothesis probability of about 11\% -- the
evidence is thus merely suggestive for metal rich clusters being more
likely to contain black holes than metal poor clusters.

Two additional candidate globular cluster black holes have been
suggested which do not exceed (or which only marginally exceed) the
Eddington luminosity for a neutron star (Barnard et al. 2008; Barnard
\& Kolb 2009) -- Bo~45 and Bo~144.  These sources have X-ray spectra
dominated by hard power law components (reasonably well-fit with
$\Gamma\approx1.5$ power laws), and X-ray luminosities which exceed
the few percent of $L_{EDD}$ at which such states are normally seen,
if the sources are neutron star accretors.  The case for these objects
being accreting black holes, while suggestive, is less secure than for
the brighter systems discussed above.  Both are redder in $r-i$ than
the median clusters in M~31 (using the confirmed old cluster sample
and SDSS photometry from Peacock et al. 2010).  Adding them into the
sample and performing a KS test gives a null hypothesis probability of
5\% that the clusters hosting black holes are more metal rich than the
cluster sample as a whole.

We can also test whether the clusters containing black hole candidates
are more luminous than the cluster population as a whole.  A slightly
more complicated procedure must be used here.  For NGC~1399A, 25/554
clusters are brighter; for the NGC~3379 BH source, 15/61 are brighter;
for NGC~4472B, 92/928 are brighter.  The other two clusters, 
NGC~1399B and NGC~4472A, are not covered by the HST cluster catalogs.
For these two clusters, we determine where their magnitudes fall in
the HST catalogs, rather than comparing with the ground-based
catalogs.  The ground-based catalog of NGC~1399 from Dirsch et
al. (2004) is especially biased in luminosity, since it represents a
spectroscopically selected sample of clusters. This sample is in $C$
and $R$, so we use the metallicity-color conversions from Smits et
al. to convert from $R$ to $I$, so that the magnitude can be compared
with the HST catalog for NGC~1399.  We thus obtain an estimate of
$I=21.3$ for NGC~1399B, brighter than all but 89/554 clusters in the
HST catalog.  The NGC~4472 brightness is directly comparable to the
existing HST catalog, and, the cluster is brighter than all but 25/928
of the NGC~4472 clusters.  Applying the same sort of KS test as was
done for color, a 2.8\% null hypothesis probability results that the
clusters are more luminous than the cluster population as a whole.
There may be a slight bias in favor of this hypothesis, given that
spectroscopic confirmation was obtained for most of these clusters
before papers were published, and that in several cases, the spectra
were already existing in the archives.  Both of the M~31 clusters
suspected to contain black holes are also considerably more massive
than the mean for M~31.

It has already been well established that the clusters containing
X-ray sources in general are more massive (Verbunt 1987) and redder
than a randomly selected sample of clusters would be (see e.g. Silk \&
Arons 1975 for the first suggestion of this effect; Bellazzini et
al. 1995 and Kundu et al. 2002 for the first strong observational
evidence for it).  In recent years, it has become clear that clusters
with higher collision rates are more likely to have bright X-ray
sources, even after accounting for the fact that such clusters are
also more massive (e.g. Jord\'an et al. 2007; Peacock et al. 2009,
2010b).  At the present time, there are not yet high enough quality
data from HST to estimate King model parameters for these clusters,
but obtaining such data would be of great interest.  The same recent
theoretical work that demonstrates that many globular clusters will
retain substantial fractions of their stellar mass black holes also
shows that the core radii of clusters with black holes can be enlarged
(Mackey et al. 2007).

\section{Summary}
We have reported the detection of a new globular cluster black hole
candidate, confirmed to be a black hole rather than a collection of
neutron stars by its strong variability.  The source is the fifth
object with this type of convincing evidence of its black hole nature.
Even with this small population of objects, it now seems likely, but
not conclusively demonstrable, that the formation of X-ray sources
with black hole accretors is favored in red clusters, and it is clear
that luminous globular clusters are more likely to host black hole
accretors.  Given the suggestions that black holes should have strong
effects on the dynamical evolution of globular clusters, estimation of
the King model parameters of these clusters would be especially
valuable.

\begin{table*}
\begin{tabular}{lccccccccc}
\hline Source & RA& Dec& $M_V$& Color & Metallicity & References\\
\hline 
NGC~4472A& 12 29 39.7& +7 53 33& V=20.99& $B-R=1.06$ &-1.7 & Maccarone et al. 2007;RZ01\\ 
NGC~4472B& 12 29 40.5& +7 57 47& V=21.77 &$V-I=1.34$& +0.5 & this paper; MKZ03\\ NGC~1399A& 3 38 31.8&-35 26 04 & I=21.0& B-I=2.25& +0.5& Irwin et al. 2010; Kundu et al. 2007\\
NGC~1399B&3 38 31.7 &-35 30 59.21 &R=22.02 &C-R=2.04 &+0.2 & Shih et al. 2010, in submission; Dirsch et al. 2004\\ 
NGC~3379&10 47 52.7 &+12 33 38.0 &V=21.88 &V-I=1.13&-0.5 & Brassington et al. 2008,2010; Kundu et al. 2007\\

\end{tabular}
\caption{The strong candidates for being globular cluster black holes.
The metallicities in the table are the color-metallicity relation of
Lee et al. for the NGC~1399 source, and the relations of Smits et
al. (2006) for the other sources.  The extreme red colors of NGC~4472B
and the NGC~1399 source may take them outside the range where the
linear metallicity-color correlations are most reliable, but it is
clear that these are among the reddest and hence most metal rich
clusters in their respective systems.
}
\label{clusters}
\end{table*}

\section{Acknowledgments} 
TJM thanks the European Union for support under FP7 grant 215212:
Black Hole Universe.  AK thanks NASA for support under Chandra grant
GO0-11111A and HST archival program HST-AR-11264.  SEZ thanks NASA for
support under grants NNX08AJ60G and Chandra GO0-11105X.  This research
is supported in part by an NSF Faculty Early Career Development
(CAREER) award (AST-0847109) to KLR.

\label{lastpage}

\end{document}